\newcommand{\citeAuthorAndIndex}[1]{\citeauthor{#1} (\citeyear{#1})~\cite{#1}}

\documentclass[sigconf]{acmart}

\AtBeginDocument{%
  }

\setcopyright{acmlicensed}
\copyrightyear{2024}
\acmYear{2024}
\acmDOI{XXXXXXX.XXXXXXX}

\acmConference[CIKM]{Make sure to enter the correct
  conference title from your rights confirmation email}{October 21--25, 2024}{Boise, Idaho, USA}
\usepackage{booktabs, multirow} % for borders and merged ranges
\begin{document}

\title{Personalization of Dataset Retrieval Results using a Metadata-based Data Valuation Method}

\author{Malick Ebiele}
\email{malick.ebiele@adaptcentre.ie}
\orcid{0000-0001-5019-6839}
\affiliation{%
  \institution{University College Dublin}
  \city{Dublin}
  % \state{Co. Dublin}
  \country{Ireland}
}

\author{Malika Bendechache}
\email{malika.bendechache@universityofgalway.ie}
\affiliation{%
  \institution{University of Galway}
  \city{Galway}
  \country{Ireland}}

\author{Eamonn Clinton}
\email{Eamonn.Clinton@tailte.ie}
\affiliation{%
  \institution{Tailte \'Eireann}
  \city{Dublin}
  \country{Ireland}
}

\author{Rob Brennan}
\email{rob.brennan@ucd.ie}
\affiliation{%
 \institution{University College Dublin}
 \city{Dublin}
 \country{Ireland}}

\renewcommand{\shortauthors}{None et al.}

\begin{abstract}
In this paper, we propose a novel data valuation method for a  Dataset Retrieval (DR) use case in Ireland’s National mapping agency.  To the best of our knowledge, data valuation has not yet been applied to Dataset Retrieval. By leveraging metadata and a user’s preferences, we estimate the personal value of each dataset to facilitate dataset retrieval and filtering. We then validated the data value-based ranking against the stakeholders’ ranking of the datasets.  The proposed data valuation method and use case demonstrated that data valuation is promising for dataset retrieval. 
For instance, the outperforming dataset retrieval based on our approach obtained 0.8207 in terms of NDCG@5 (the truncated Normalized Discounted Cumulative Gain at 5).
% For instance,  personalized dataset retrieval based on data valuation methods can improve the performance of such systems. 
This study is unique in its exploration of a data valuation-based approach to dataset retrieval and stands out because, unlike most existing methods, our approach is validated using the stakeholders’ ranking of the datasets.
\end{abstract}





\maketitle

\section{Introduction}
Given rapidly rising data volumes, knowing which data to keep and which to discard has become an essential task. Data valuation has emerged as a promising approach to tackle this problem \cite{even_value-driven_2005}. 
% Data valuation is an area of study encompassing but not limited to data quality, machine learning, applied energy, and information economics (infonomics). 
The primary focus of data valuation research is the development of methodologies for determining the value of data \cite{wang_principled_2020, wang_data_2021, laney_infonomics_2017, turczyk_method_2007, khokhlov_what_2020,  qiu_evaluation_2017}. 
	
 Data valuation methods have been applied to data management, machine learning, system security, and energy \cite{wang_principled_2020, turczyk_method_2007, khokhlov_what_2020, wang_data_2021}, to name a few. No previous attempts to apply data valuation to dataset retrieval. Dataset retrieval is a specialization of information retrieval with the difference that instead of retrieving relevant documents the engine returns a list of relevant datasets \cite{kunze_dataset_2013}. Dataset retrieval systems will return relevant datasets according to a given query. However, they do not take into account the preferences of the user in terms of metadata. Some dataset retrieval softwares allow users to sort the results by each metadata separately like creation date, usage, and last update. But none of them allow the users to sort the results by a combination of those metadata. In this paper, we proposed a metadata-based data valuation method that will allow users to sort dataset retrieval results using a combination of metadata.

 Many of the existing data valuation approaches are subjective due to the subjective nature of some dimensions that characterize data value \cite{panetto_challenges_2018} or the subjective weighting techniques (in the case of weighted averaging or summing) \cite{odu_weighting_2019, deng_electricity_2023}. Subjective metrics of data value dimensions or weighting techniques can only be defined by individual users or experts based on their personal views, experiences, and backgrounds. These are opposed to objective metrics that can be determined precisely on the basis of a detailed analysis of the data or extracted from the data infrastructure \cite{bodendorf_scientific_2022}. This makes it challenging to develop a fully objective data valuation model. We believe that instead of generalizing subjective metrics and weighting techniques, it would be better to attempt to develop personalized data valuation models. 
	
Choosing a suitable weighting technique is an additional challenge for weighted approaches. For instance, usage-over-time is one of the first data valuation methods and developed a weighting technique based on recency \cite{chen_information_2005}. The recency-based weighting technique is per objective. However, the choice of using recency-based as a weighting technique is subjective or the choice of assigning high or low weights to the more recency metadata is also subjective. In our case, the desired weighting technique should be subjective and straightforward for the stakeholders to interact with. 

The research question is: To what extent can metadata-based data valuation methods improve the results of dataset retrieval systems in terms of ranking performance?
	
To answer the research question above, we designed and implemented a metadata-based data valuation method and applied it to a dataset retrieval use case for a National Mapping Agency. The goal is to make retrieval in a data catalog more effective by correctly identifying the datasets that are more valuable to customers by taking into account the customers' dataset preferences. 

The contributions of this paper are as follows:

\begin{itemize}
    \item First attempt to apply a metadata-based data valuation method to dataset retrieval.
    \item Proposed a personalized and interactive data valuation method. Extant methods are mainly subjective approaches.
\end{itemize}
 
The remainder of this paper is structured as follows. Section 2 gives a description of the use case. Section 3 describes the related work. Our proposed metadata-based data valuation method is explained in Section 4. Section 5 explains our experimental design. In Section 6, the experimental results are shown and discussed. Finally, the conclusion and future work are presented in Section 7.

% \section{Use Case Description}
%% This section has been commented out because it can be used to identify the authors' ethnicity, country of origin, location, etc.
% This data valuation project is part of an ongoing collaboration between researchers from University College Dublin (UCD) and Tailte Éireann (TE). Tailte Éireann (TE) is Ireland's state agency for property registrations, property valuation and national mapping services. It was established on 1 March 2023 from a merger of the Property Registration Authority (PRA), the Valuation Office (VO) and Ordnance Survey Ireland (OSI). The end goal of this collaboration is to design and implement a data valuation method for TE datasets from the customer's perspective. The data valuation method should take into account the customers' preferences in terms of metadata. At this stage, the goal is to design and implement a proof of concept. 

\section{Related Work}
This section covers different approaches to metadata-based data valuation methods. It highlights their challenges especially challenges related to weighted average approaches because the approach proposed in this paper falls into that category.

\textbf{Overview of data valuation methods}.\citeAuthorAndIndex{wijnhoven_value-based_2014} used file age and last modification time metadata to train machine learning models to classify files as waste (or nonwaste). Random Forest achieved a performance of  84\% accuracy using 300 data points (20\% for training). The main problem with this approach is the scarcity of labeled data. To achieve better performance, more data will be needed. The main reason Random Forest achieved  84\% accuracy is because the classification problem is binary classification with only two dependent variables which is quite straightforward. In many cases, there are more than two dependent variables and classes. Another approach is Fuzzy expert system. \citeAuthorAndIndex{hanratty_enhancing_2013} used Fuzzy Associative Memory (FAM) with three inputs: source reliability, information content, and timeliness to calculate the value of information [data] in a military application context. There are two options here: either subjectively choosing the membership functions of the Fuzzy system or fine-tuning using data to objectively select the membership functions. This is not the direction we would like to follow in this paper because our first choice will be fine-tuning the Fuzzy system using data which will require a lot more data than we can currently collect.
%Fuzzy systems are known to be good at approximating reasoning where information is uncertain, incomplete, imprecise, and vague \cite{hanratty_enhancing_2013}. 
Probability and Statistics and Matrix multiplication are other approaches to metadata-based data valuation.  Probability and Statistics are applied to estimate the probability of further use of a dataset \cite{turczyk_method_2007}. As to matrix multiplication, it is chosen in the context of two-level indicators data valuation \cite{chen_value_2011, fang_information_2016}. These approaches are not relevant because our proposed approach is not a two-level data valuation method and we are also not trying to predict the probability of future use of the datasets.

\textbf{Weighted average data valuation methods}. There were also previous attempts to calculate the data value using weighted averaging \cite{chen_information_2005, ma_mdv_2019, qiu_evaluation_2017}. For instance, usage-over-time is one of the first data valuation methods used and it estimates data value with a weighted averaging approach \cite{chen_information_2005}. It consists of splitting the usage data into a series of time slots, assigning a weight to each time slot, and then computing the data value using the weighted average. The weights are the normalized recency weights. The more recent time slots are assigned the higher weights \cite{chen_information_2005}. \citeAuthorAndIndex{ma_mdv_2019} extended the usage-over-time model by adding the age and size dimensions. Their  Multi-Factors Data Valuation Method ($MDV$) is a trade-off between dynamic and static data value. The dynamic data value is the usage-over-time model of Chen. The static data value is the weighted average of the normalized age and size. The weights of the age and size are assigned subjectively by experts. 

\citeAuthorAndIndex{qiu_evaluation_2017} used the Analytical Hierarchy Process (AHP) which is a different weighting approach. AHP requires a subjective rating of the input dimensions in pairs. These pairwise comparisons are then arranged in a matrix, from which a final weighting of the dimensions will be calculated. AHP is technically straightforward to implement and more importantly allows to assess the transitivity consistency of the pairwise comparisons matrix by assigning a consistency score to it. However, experts are still needed for the pairwise rating of the input dimensions. \citeAuthorAndIndex{qiu_evaluation_2017} use the measure of 6 dimensions in their model. For more details on the dimensions used, please refer to \cite{qiu_evaluation_2017}. 
% Those dimensions are: The size of the data ($S$), the access interval ($T$), the data read and write frequency ($F$), the number of visits ($C$), the contents of the file ($D$, which is the sum of the ratio of the total number of words in the storage system n and the number of occurrences $m_i$ of the word $w_i$  in the storage system), and the potential value of the data ($V$, which is the measure of the number of potential users by calculating the similarity between users' access data and no access data).

The challenge of weighted approaches is the weighting technique. In our case, the desired weighting technique should be straightforward for the stakeholders to interact with. The weighting approach used in this paper is detailed in Section 4.1.

To the best of our knowledge, the data valuation approach proposed in this study and the application of data value to dataset retrieval is unique. Also, none of the studies described above validated the outputs of their data valuation approaches. Our approach is validated using data collected from stakeholders.

\section{Proposed Data Valuation Method}
\subsection{Weight Determination}
Analytic Hierarchy Process (AHP) was our first choice because of its sound mathematical basis \cite{saaty_analytic_1987}. However, it was challenging to interact with, as instead of assigning a weight to each metadata or dimension, a pairwise comparison of the dimensions is needed \cite{saaty_analytic_1987}. E.g. usage is twice as important as creation date, usage is 5 times more important than number of spatial objects, or usage is twice less important than utility. This exercise was difficult for the stakeholders who participated in the experiments. They confessed to being more comfortable with a rating-like weighting approach e.g. 1 to 5 websites or products rating mechanism. Also, AHP assumes that preferences are transitive and has a transitivity consistency test. \citeAuthorAndIndex{saaty_analytic_1987} advise to discard the current weights deduced from the pairwise comparisons if the consistency ratio is greater than 0.1. There are previous studies showing that preferences are not always transitive \cite{alos-ferrer_identifying_2023, alos-ferrer_choice_2021, gendin_why_1996, fishburn_nontransitive_1991}. \citeAuthorAndIndex{alos-ferrer_identifying_2023} shows using two preference datasets that no matter the initial assumptions, even when the preferences are supposed to be transitive, a maximum of 27.45\% of individual preferences are non-transitive. We believe that assuming that all preferences are transitive implies ignoring some individual preferences. Therefore, we used a slider from 0 to 10 (with a step of 1) as the weights determination technique; the presence of a zero rating allows the individual to discard a particular metadata as irrelevant to the use case under consideration. This approach seems to be straightforward and inclusive after testing it during the interviews with the stakeholders. The only constraint in our weighting approach is that at least one of the provided weights should be non-zero.

\subsection{Data Value Calculation}
As the collected metadata values have different scales, they must be normalized. The weights also must be normalized. For usage and number of objects metadata, the values must be divided by the maximum value. Utility values have been normalized by dividing the values by 100 (the maximum possible utility value). As to the creation date, we applied the probabilistic approach of calculating data currency with a decline rate of 20\%. This approach was proposed by \citeAuthorAndIndex{heinrich_assessing_2011} and the data currency $Q_{Curr.}(\omega, A)$ formula is shown in the Equation \ref{eq:01} below.

\begin{equation}\label{eq:01}
    Q_{Curr.}(\omega, A) := exp(-decline(A) \cdot age(\omega, A))
\end{equation}

For the weights, the weight of each metadata has been divided by the sum of the weights of all four metadata per stakeholder. The data value is then the weighted average of the metadata values using the Equation \ref{eq:02} below. 
\begin{equation}\label{eq:02}
	V(d_i) = w_{\text{U}} \times \text{U}_i + w_{\text{S}} \times \text{S}_i + w_{\text{C}} \times \text{C}_i + w_{\text{O}} \times \text{O}_i
\end{equation}
Where $w_{\{\text{U, S, C, O}\}}$ in [0,1] are the weights and $V(d_i)$ in ]0,1] the data value. U, S, C, and O stand for Utility, uSage, Creation date, and number of Objects, respectively.

\section{Experimental Design}
Figure \ref{fig:experimental_design} below shows the flowchart of our experimental design.  The experiments consist of three main steps: Data collection, Data value calculation, and Analysis and conclusions.

%\begin{itemize}
    \vspace{2mm}
    \noindent \textbf{Step 1:} Data collection. It has two main substeps: metadata collection and interview of stakeholders. The stakeholders included in this study are managers within the mapping agency with data management responsibilities for at least 3 years each.
    %\item Step 1: Data collection. It has two main substeps: metadata collection and interview of stakeholders.
    \begin{itemize}
        \item Step 1.1: Metadata collection. This consists of extracting metadata from the dataset repository system. For this use case, only three metadata types have been extracted:  creation date, number of spatial objects (a domain-relevant measure of data volume and information content), and usage. We also collected, using spreadsheets, an estimation of the utility of the datasets from four stakeholders who were domain experts.
        \item Step 1.2: Stakeholder interview. The main goal of each interview (15-20 minutes) was to get the stakeholders to assign weights to each of the metadata and their ideal dataset ranking based on a data retrieval scenario. A slider from 0 to 10 (with a step of 1) is used to assign the weight to each value metadata. Three stakeholders participated in the interview with only one of them who also provided an estimation of the utility of each dataset.
    \end{itemize}

%    \item Step 2: Data value calculation. So far, four metadata have been collected on 15 datasets and three sets of weights from three stakeholders. The data valuation process described in Section 4.2 is used to estimate the value of each dataset.
    \noindent \textbf{Step 2:} Data value calculation. Four metadata have been collected on 15 datasets and three sets of weights from three stakeholders. The data valuation process described in Section 4.2 is used to estimate the value of each dataset for each stakeholder.

    %\item Step 3: Analysis and conclusions. Here, the data value-based dataset ranking is compared with the provided datasets ranking provided by the stakeholders. The main objective is to check if both rankings match. The findings are discussed in detail in the following Section.
   \vspace{2mm}
   \noindent \textbf{Step 3:} Analysis and conclusions. The data value-based dataset ranking is compared with the dataset ranking provided by the stakeholders. The main objective is to check to what extent both rankings match. The findings are discussed in the following Section.
%\end{itemize}

\begin{figure*}[h]
  \centering
  \includegraphics[width=0.85\textwidth]{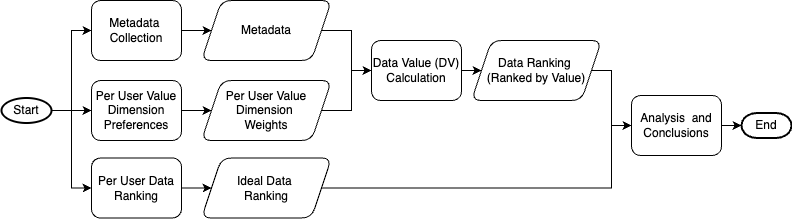}
  \caption{Experimental design for personalized metadata-based data valuation.}
  % \Description{}
  \label{fig:experimental_design}
\end{figure*} 

\section{Experimental Results}

% The 15 datasets have been sorted alphabetically and assigned an identification number from 1 to 15. 
Table \ref{tab:weights} shows the weights provided by each stakeholder. Stakeholder 2 (SH2) provided an invalid set of weights (all of the weights are zero) because all of the metadata selected for this case study was invalid for them. This shows the practicality of giving the possibility of users or experts to assign zero weights.

The personal data value was calculated for each dataset using the valid weights provided by stakeholders SH1 and SH3 and eqt. \ref{eq:02}. The datasets were then ranked by data value. The resulting calculated personalized ranking was then compared to the ideal ranking provided by the stakeholder. The ideal ranking of each stakeholder has also been compared to special cases where the datasets are ranked using each metadata independently (univariate ranking).  

The Normalized Discounted Cumulative Gain (NDCG) is chosen to evaluate the performance of the data value-based rankings. There are two main reasons for choosing NDCG. Firstly, NDCG is widely used and involves a discount function over the rank while many other measures uniformly weight all positions. This feature is important for our use case but also for search engines as users care about top-ranked documents much more than others; the truncated NDCG also known as NDCG@k, which only focuses on the top k elements, is preferred for our use case. Secondly, NDCG is proven to be a good ranking measure from a learning-to-rank point of view \cite{wang_theoretical_2013}. NDCG ranges between 0 and 1, with 1 being the optimal performance. We used the scikit-learn implementation of NDCG with the default parameters\footnote{\url{https://scikit-learn.org/stable/modules/generated/sklearn.metrics.ndcg_score.html}}.  

\subsection{Evaluation Results}
Table \ref{tab:evaluation_results} presents the evaluation results. The performance of the different data value-based rankings is expressed in terms of NDCG and  NDCG@5. 

NDCG and NDCG@5 are calculated per stakeholder and data value calculation technique (weighted average, simple average and univariate). Simple average and univariate data valuation are special cases of weighted approaches where all the weights are equal and where only one weight is non-zero, respectively. 

We highlighted the best NDCG and NDCG@5 per stakeholder and data value calculation technique.

As Usage metadata has been collected monthly from January 2017 to January 2023, we calculated the total and average usage and used them as independent metadata. For Utility, only experimental stakeholder 1 provided a Utility estimate and weights of the metadata. Therefore, we included their Utility estimate as metadata. We also computed the average of the Utility metadata and used it as another metadata. 

\subsection{Discussion}
NDCG and the univariate rankings help us to identify which metadata is the best determinant in each of the three stakeholders' ideal ranking of the datasets. The most determinant metadata for stakeholder 1 are Creation date and Usage, respectively. For stakeholders 2 and 3, the most determinant metadata are Utility and Creation date, on one hand, and Utility and Number of spatial objects, on the other, respectively.

NDCG@5 measures the dataset retrieval performance of the weighted average data valuation method and its variants. For stakeholder 1, the outperforming ranking method is the Creation date-based univariate and the second best is the weighted average using their provided Utility estimate. For stakeholder 2, the outperforming ranking method is the simple average using average Usage and the second best is the Creation date-based univariate. For stakeholder 3, the outperforming ranking methods, with a tie, are the average Utility-based and stakeholder 1 provided Utility-based univariate and the third best is the Creation date-based univariate.

\begin{table}
  \caption{Dataset value dimension (metadata field) weights provided by stakeholders}
  \label{tab:weights}
  \begin{tabular}{ccccc}
    \toprule
    \begin{tabular}[c]{@{}c@{}}Stakeholders\\ (SH) \end{tabular} & Utility & Creation Date & \begin{tabular}[c]{@{}c@{}} \#Spatial Objects\end{tabular} & Usage\\
    \midrule
    SH1 &  8  & 10 & 8 & 5 \\
    SH2 & 0  & 0 & 0 & 0 \\
    SH3  & 7 & 9 & 9 & 4 \\
  \bottomrule
\end{tabular}
\end{table}

%If the table is too wide, replace \begin{table}[!htp]...\end{table} with
%\begin{adjustwidth}{-2.5 cm}{-2.5 cm}\centering\begin{threeparttable}[!htb]...\end{threeparttable}\end{adjustwidth}
\begin{table}[!htp]\centering
\caption{Evaluation results}\label{tab:evaluation_results}
\scriptsize
\begin{tabular}{llllrr}
\toprule
& & &NDCG &NDCG@5 \\
\midrule
\multirow{12}{*}{Stakeholder 1} &\multirow{3}{*}{Weighted Average Raking} &Total Usage &0.7522 &0.3988 \\
& &Average Usage &0.7481 &0.3868 \\
& &Provided Utility &\textbf{0.9121} &\textbf{0.7469} \\
\cmidrule{2-5}
&\multirow{3}{*}{Simple Average Raking} &Total Usage &\textbf{0.8057} &\textbf{0.5853} \\
& &Average Usage &0.7766 &0.4662 \\
& &Provided Utility &0.7771 &0.4564 \\
\cmidrule{2-5}
&\multirow{6}{*}{Univariate Raking} &Stakeholder 1 Provided Utility &0.8062 &0.6288 \\
& &Average Utility &0.8047 &0.6288 \\
& &Number of Spatial Objects &0.8002 &0.5331 \\
& &Creation Date &\textbf{0.8984} &\textbf{0.8984} \\
& &Total Usage &0.8881 &0.6869 \\
& &Average Usage &0.8870 &0.6869 \\
\hline
\multirow{8}{*}{Stakeholder 2} &\multirow{2}{*}{Simple Average Raking} &Total Usage &0.9159 &0.7557 \\
& &Average Usage &\textbf{0.9303} &\textbf{0.8207} \\
\cmidrule{2-5}
&\multirow{6}{*}{Univariate Raking} &Stakeholder 1 Provided Utility &\textbf{0.9298} &0.7694 \\
& &Average Utility &0.9195 &0.7694 \\
& &Number of Spatial Objects &0.7797 &0.5095 \\
& &Creation Date &0.8150 &\textbf{0.8150} \\
& &Total Usage &0.7190 &0.3652 \\
& &Average Usage &0.7241 &0.3652 \\
\hline
\multirow{10}{*}{Stakeholder 3} &\multirow{2}{*}{Weighted Average Raking} &Total Usage &\textbf{0.8155} &\textbf{0.4756} \\
& &Average Usage &\textbf{0.8155} &\textbf{0.4756} \\
\cmidrule{2-5}
&\multirow{2}{*}{Simple Average Raking} &Total Usage &\textbf{0.8450} &\textbf{0.6153} \\
& &Average Usage &0.8148 &0.4853 \\
\cmidrule{2-5}
&\multirow{6}{*}{Univariate Raking} &Stakeholder 1 Provided Utility &0.9331 &\textbf{0.9187} \\
& &Average Utility &\textbf{0.9423} &\textbf{0.9187} \\
& &Number of Spatial Objects &0.8577 &0.5680 \\
& &Creation Date &0.8505 &0.8505 \\
& &Total Usage &0.8116 &0.5650 \\
& &Average Usage &0.8053 &0.5650 \\
\bottomrule
\end{tabular}
\end{table}

\section{Conclusion}
This paper presents a weighted average data valuation method. The proposed method is then applied to a dataset retrieval use case. The best two (weighted or simple) averaging performances in terms of NDCG@5 are stakeholder 2 simple average ranking using average Usage at 0.8207 and stakeholder 1 weighted average ranking using their own provided Utility at 0.7469. This shows that the proposed approach could successfully be applied to dataset retrieval. However, more work is needed to improve the performance of the proposed method.

There are many directions that this research can follow for further explorations. For instance, stakeholders, users, or experts can be asked to provide an ideal ranking of the datasets given a set of metadata. Then, use NDCG and the univariate rankings to statistically analyse the significance of the weights of each metadata. This could be a new objective weighting technique. Another direction could be to compute data value using all the possible weight distributions (weighted and simple average, univariate, bivariate, trivariate, and so on). A third direction could be combining the previous two.

% \begin{acks}
% This research has received funding from the ADAPT Centre for Digital Content Technology, funded under the SFI Research Centres Programme (Grant 13/RC/2106\_P2), co-funded by the European Regional Development Fund. For the purpose of Open Access, the author has applied a CC BY public copyright licence to any Author Accepted Manuscript version arising from this submission.
% \end{acks}

\bibliographystyle{ACM-Reference-Format}
\bibliography{references}


\begin{thebibliography}{25}

%%% the Web address.

\ifx \showCODEN    \undefined \def \showCODEN     #1{\unskip}     \fi
\ifx \showDOI      \undefined \def \showDOI       #1{#1}\fi
\ifx \showISBNx    \undefined \def \showISBNx     #1{\unskip}     \fi
\ifx \showISBNxiii \undefined \def \showISBNxiii  #1{\unskip}     \fi
\ifx \showISSN     \undefined \def \showISSN      #1{\unskip}     \fi
\ifx \showLCCN     \undefined \def \showLCCN      #1{\unskip}     \fi
\ifx \shownote     \undefined \def \shownote      #1{#1}          \fi
\ifx \showarticletitle \undefined \def \showarticletitle #1{#1}   \fi
\ifx \showURL      \undefined \def \showURL       {\relax}        \fi
% invisible to TeX
\providecommand\bibfield[2]{#2}
\providecommand\bibinfo[2]{#2}
\providecommand\natexlab[1]{#1}
\providecommand\showeprint[2][]{arXiv:#2}

\bibitem[Alós-Ferrer et~al\mbox{.}(2023)]%
        {alos-ferrer_identifying_2023}
\bibfield{author}{\bibinfo{person}{Carlos Alós-Ferrer}, \bibinfo{person}{Ernst Fehr}, {and} \bibinfo{person}{Michele Garagnani}.} \bibinfo{year}{2023}\natexlab{}.
\newblock \showarticletitle{Identifying nontransitive preferences}.
\newblock  (\bibinfo{date}{Jan.} \bibinfo{year}{2023}).
\newblock
\urldef\tempurl%
\url{https://doi.org/10.5167/UZH-219280}
\showDOI{\tempurl}
\newblock
\shownote{Publisher: [object Object]}.


\bibitem[Alós-Ferrer and Garagnani(2021)]%
        {alos-ferrer_choice_2021}
\bibfield{author}{\bibinfo{person}{Carlos Alós-Ferrer} {and} \bibinfo{person}{Michele Garagnani}.} \bibinfo{year}{2021}\natexlab{}.
\newblock \showarticletitle{Choice consistency and strength of preference}.
\newblock \bibinfo{journal}{\emph{Economics Letters}}  \bibinfo{volume}{198} (\bibinfo{date}{Jan.} \bibinfo{year}{2021}), \bibinfo{pages}{109672}.
\newblock
\showISSN{0165-1765}
\urldef\tempurl%
\url{https://doi.org/10.1016/j.econlet.2020.109672}
\showDOI{\tempurl}


\bibitem[Attard and Brennan(2018)]%
        {panetto_challenges_2018}
\bibfield{author}{\bibinfo{person}{Judie Attard} {and} \bibinfo{person}{Rob Brennan}.} \bibinfo{year}{2018}\natexlab{}.
\newblock \showarticletitle{Challenges in {Value}-{Driven} {Data} {Governance}}.
\newblock In \bibinfo{booktitle}{\emph{On the {Move} to {Meaningful} {Internet} {Systems}. {OTM} 2018 {Conferences}}}, \bibfield{editor}{\bibinfo{person}{Hervé Panetto}, \bibinfo{person}{Christophe Debruyne}, \bibinfo{person}{Henderik~A. Proper}, \bibinfo{person}{Claudio~Agostino Ardagna}, \bibinfo{person}{Dumitru Roman}, {and} \bibinfo{person}{Robert Meersman}} (Eds.). Vol.~\bibinfo{volume}{11230}. \bibinfo{publisher}{Springer International Publishing}, \bibinfo{address}{Cham}, \bibinfo{pages}{546--554}.
\newblock
\showISBNx{978-3-030-02670-7 978-3-030-02671-4}
\urldef\tempurl%
\url{https://doi.org/10.1007/978-3-030-02671-4_33}
\showDOI{\tempurl}
\newblock
\shownote{Series Title: Lecture Notes in Computer Science}.


\bibitem[Bodendorf et~al\mbox{.}(2022)]%
        {bodendorf_scientific_2022}
\bibfield{author}{\bibinfo{person}{Frank Bodendorf}, \bibinfo{person}{Klaus Dehmel}, {and} \bibinfo{person}{Joerg Franke}.} \bibinfo{year}{2022}\natexlab{}.
\newblock \bibinfo{booktitle}{\emph{Scientific {Approaches} and {Methodology} to {Determine} the {Value} of {Data} as an {Asset} and {Use} {Case} in the {Automotive} {Industry}}}.
\newblock
\showISBNx{978-0-9981331-5-7}
\urldef\tempurl%
\url{http://hdl.handle.net/10125/80023}
\showURL{%
\tempurl}


\bibitem[Chen(2005)]%
        {chen_information_2005}
\bibfield{author}{\bibinfo{person}{Ying Chen}.} \bibinfo{year}{2005}\natexlab{}.
\newblock \showarticletitle{Information valuation for information lifecycle management}. In \bibinfo{booktitle}{\emph{Second {International} {Conference} on {Autonomic} {Computing} ({ICAC}'05)}}. \bibinfo{publisher}{IEEE}, \bibinfo{pages}{135--146}.
\newblock


\bibitem[Chen et~al\mbox{.}(2011)]%
        {chen_value_2011}
\bibfield{author}{\bibinfo{person}{Yanshou Chen}, \bibinfo{person}{Xiaohui Wu}, {and} \bibinfo{person}{Xuejuan Yang}.} \bibinfo{year}{2011}\natexlab{}.
\newblock \showarticletitle{Value evaluation of enterprise information based on grid-fuzzy borda number analytical method}. In \bibinfo{booktitle}{\emph{2011 {International} {Conference} on {E}-{Business} and {E}-{Government} ({ICEE})}}. \bibinfo{pages}{1--5}.
\newblock
\urldef\tempurl%
\url{https://doi.org/10.1109/ICEBEG.2011.5881899}
\showDOI{\tempurl}


\bibitem[Deng et~al\mbox{.}(2023)]%
        {deng_electricity_2023}
\bibfield{author}{\bibinfo{person}{Yong Deng}, \bibinfo{person}{Xiaoming Li}, \bibinfo{person}{Xiang Yu}, \bibinfo{person}{Zhening Fu}, \bibinfo{person}{Hancheng Chen}, \bibinfo{person}{Jingyu Xie}, {and} \bibinfo{person}{Danhong Xie}.} \bibinfo{year}{2023}\natexlab{}.
\newblock \showarticletitle{Electricity {Data} {Valuation} {Considering} {Attribute} {Weights}}. In \bibinfo{booktitle}{\emph{2023 3rd {International} {Conference} on {Intelligent} {Power} and {Systems} ({ICIPS})}}. \bibinfo{pages}{788--794}.
\newblock
\urldef\tempurl%
\url{https://doi.org/10.1109/ICIPS59254.2023.10404587}
\showDOI{\tempurl}


\bibitem[Even and Shankaranarayanan(2005)]%
        {even_value-driven_2005}
\bibfield{author}{\bibinfo{person}{Adir Even} {and} \bibinfo{person}{Ganesan Shankaranarayanan}.} \bibinfo{year}{2005}\natexlab{}.
\newblock \showarticletitle{Value-{Driven} {Data} {Quality} {Assessment}.}. In \bibinfo{booktitle}{\emph{{ICIQ}}}.
\newblock


\bibitem[Fang(2016)]%
        {fang_information_2016}
\bibfield{author}{\bibinfo{person}{Junpeng Fang}.} \bibinfo{year}{2016}\natexlab{}.
\newblock \showarticletitle{Information {Value} {Evaluation} {Index} {System}}. In \bibinfo{booktitle}{\emph{{PROCEEDINGS} {OF} {THE} 2016 {2ND} {WORKSHOP} {ON} {ADVANCED} {RESEARCH} {AND} {TECHNOLOGY} {IN} {INDUSTRY} {APPLICATIONS}}}, \bibfield{editor}{\bibinfo{person}{Z.~Zeng} {and} \bibinfo{person}{X.~Bai}} (Eds.), Vol.~\bibinfo{volume}{81}. \bibinfo{publisher}{Atlantis Press}, \bibinfo{address}{Paris}, \bibinfo{pages}{1045--1048}.
\newblock
\showISBNx{978-94-6252-195-7}
\urldef\tempurl%
\url{https://www.webofscience.com/wos/woscc/summary/fceb06eb-e340-45d9-be08-1961b34d32d1-cf48e49b/relevance/1}
\showURL{%
\tempurl}
\newblock
\shownote{ISSN: 2352-5401 Num Pages: 4 Series Title: AER-Advances in Engineering Research Web of Science ID: WOS:000388364600221}.


\bibitem[Fishburn(1991)]%
        {fishburn_nontransitive_1991}
\bibfield{author}{\bibinfo{person}{Peter~C. Fishburn}.} \bibinfo{year}{1991}\natexlab{}.
\newblock \showarticletitle{Nontransitive {Preferences} in {Decision} {Theory}}.
\newblock \bibinfo{journal}{\emph{Journal of Risk and Uncertainty}} \bibinfo{volume}{4}, \bibinfo{number}{2} (\bibinfo{year}{1991}), \bibinfo{pages}{113--134}.
\newblock
\showISSN{0895-5646}
\urldef\tempurl%
\url{https://www.jstor.org/stable/41760621}
\showURL{%
\tempurl}
\newblock
\shownote{Publisher: Springer}.


\bibitem[Gendin(1996)]%
        {gendin_why_1996}
\bibfield{author}{\bibinfo{person}{Sidney Gendin}.} \bibinfo{year}{1996}\natexlab{}.
\newblock \showarticletitle{Why {Preference} is {Not} {Transitive}}.
\newblock \bibinfo{journal}{\emph{The Philosophical Quarterly (1950-)}} \bibinfo{volume}{46}, \bibinfo{number}{185} (\bibinfo{year}{1996}), \bibinfo{pages}{482--488}.
\newblock
\showISSN{0031-8094}
\urldef\tempurl%
\url{https://doi.org/10.2307/2956357}
\showDOI{\tempurl}
\newblock
\shownote{Publisher: [Oxford University Press, University of St. Andrews, Scots Philosophical Association]}.


\bibitem[Hanratty et~al\mbox{.}(2013)]%
        {hanratty_enhancing_2013}
\bibfield{author}{\bibinfo{person}{Timothy~P. Hanratty}, \bibinfo{person}{Robert~J. Hammell~II}, \bibinfo{person}{Barry~A. Bodt}, \bibinfo{person}{Eric~G. Heilman}, {and} \bibinfo{person}{John~C. Dumer}.} \bibinfo{year}{2013}\natexlab{}.
\newblock \showarticletitle{Enhancing battlefield situational awareness through fuzzy-based value of information}. In \bibinfo{booktitle}{\emph{2013 46th {Hawaii} {International} {Conference} on {System} {Sciences}}}. \bibinfo{publisher}{IEEE}, \bibinfo{pages}{1402--1411}.
\newblock


\bibitem[Heinrich and Klier(2011)]%
        {heinrich_assessing_2011}
\bibfield{author}{\bibinfo{person}{Bernd Heinrich} {and} \bibinfo{person}{Mathias Klier}.} \bibinfo{year}{2011}\natexlab{}.
\newblock \showarticletitle{Assessing data currency—a probabilistic approach}.
\newblock \bibinfo{journal}{\emph{Journal of Information Science}} \bibinfo{volume}{37}, \bibinfo{number}{1} (\bibinfo{year}{2011}), \bibinfo{pages}{86--100}.
\newblock
\newblock
\shownote{Publisher: Sage Publications Sage UK: London, England}.


\bibitem[Khokhlov and Reznik(2020)]%
        {khokhlov_what_2020}
\bibfield{author}{\bibinfo{person}{Igor Khokhlov} {and} \bibinfo{person}{Leon Reznik}.} \bibinfo{year}{2020}\natexlab{}.
\newblock \showarticletitle{What is the value of data value in practical security applications}. In \bibinfo{booktitle}{\emph{2020 {IEEE} {Systems} {Security} {Symposium} ({SSS})}}. \bibinfo{publisher}{IEEE}, \bibinfo{pages}{1--8}.
\newblock


\bibitem[Kunze and Auer(2013)]%
        {kunze_dataset_2013}
\bibfield{author}{\bibinfo{person}{Sven~R. Kunze} {and} \bibinfo{person}{Sören Auer}.} \bibinfo{year}{2013}\natexlab{}.
\newblock \showarticletitle{Dataset {Retrieval}}. In \bibinfo{booktitle}{\emph{2013 {IEEE} {Seventh} {International} {Conference} on {Semantic} {Computing}}}. \bibinfo{pages}{1--8}.
\newblock
\urldef\tempurl%
\url{https://doi.org/10.1109/ICSC.2013.12}
\showDOI{\tempurl}


\bibitem[Laney(2017)]%
        {laney_infonomics_2017}
\bibfield{author}{\bibinfo{person}{Douglas~B. Laney}.} \bibinfo{year}{2017}\natexlab{}.
\newblock \bibinfo{booktitle}{\emph{Infonomics: how to monetize, manage, and measure information as an asset for competitive advantage}}.
\newblock \bibinfo{publisher}{Routledge}.
\newblock


\bibitem[Ma and Zhang(2019)]%
        {ma_mdv_2019}
\bibfield{author}{\bibinfo{person}{Xiao Ma} {and} \bibinfo{person}{Xu Zhang}.} \bibinfo{year}{2019}\natexlab{}.
\newblock \showarticletitle{{MDV}: {A} {Multi}-{Factors} {Data} {Valuation} {Method}}. In \bibinfo{booktitle}{\emph{2019 5th {International} {Conference} on {Big} {Data} {Computing} and {Communications} ({BIGCOM})}}. \bibinfo{pages}{48--53}.
\newblock
\urldef\tempurl%
\url{https://doi.org/10.1109/BIGCOM.2019.00016}
\showDOI{\tempurl}


\bibitem[Odu(2019)]%
        {odu_weighting_2019}
\bibfield{author}{\bibinfo{person}{G.~O. Odu}.} \bibinfo{year}{2019}\natexlab{}.
\newblock \showarticletitle{Weighting methods for multi-criteria decision making technique}.
\newblock \bibinfo{journal}{\emph{Journal of Applied Sciences and Environmental Management}} \bibinfo{volume}{23}, \bibinfo{number}{8} (\bibinfo{year}{2019}), \bibinfo{pages}{1449--1457}.
\newblock
\urldef\tempurl%
\url{https://www.ajol.info/index.php/jasem/article/view/189641}
\showURL{%
\tempurl}


\bibitem[Qiu et~al\mbox{.}(2017)]%
        {qiu_evaluation_2017}
\bibfield{author}{\bibinfo{person}{Shaoming Qiu}, \bibinfo{person}{Dong Zhang}, {and} \bibinfo{person}{Xiuli Du}.} \bibinfo{year}{2017}\natexlab{}.
\newblock \showarticletitle{An {Evaluation} {Method} of {Data} {Valuation} {Based} on {Analytic} {Hierarchy} {Process}}. In \bibinfo{booktitle}{\emph{2017 14th {International} {Symposium} on {Pervasive} {Systems}, {Algorithms} and {Networks} \& 2017 11th {International} {Conference} on {Frontier} of {Computer} {Science} and {Technology} \& 2017 {Third} {International} {Symposium} of {Creative} {Computing} ({ISPAN}-{FCST}-{ISCC})}}. \bibinfo{pages}{524--528}.
\newblock
\urldef\tempurl%
\url{https://doi.org/10.1109/ISPAN-FCST-ISCC.2017.21}
\showDOI{\tempurl}
\newblock
\shownote{ISSN: 2375-527X}.


\bibitem[Saaty(1987)]%
        {saaty_analytic_1987}
\bibfield{author}{\bibinfo{person}{R.~W. Saaty}.} \bibinfo{year}{1987}\natexlab{}.
\newblock \showarticletitle{The analytic hierarchy process—what it is and how it is used}.
\newblock \bibinfo{journal}{\emph{Mathematical Modelling}} \bibinfo{volume}{9}, \bibinfo{number}{3} (\bibinfo{date}{Jan.} \bibinfo{year}{1987}), \bibinfo{pages}{161--176}.
\newblock
\showISSN{0270-0255}
\urldef\tempurl%
\url{https://doi.org/10.1016/0270-0255(87)90473-8}
\showDOI{\tempurl}


\bibitem[Turczyk et~al\mbox{.}(2007)]%
        {turczyk_method_2007}
\bibfield{author}{\bibinfo{person}{Lars Turczyk}, \bibinfo{person}{Marcel Groepl}, \bibinfo{person}{Nicolas Liebau}, {and} \bibinfo{person}{Ralf Steinmetz}.} \bibinfo{year}{2007}\natexlab{}.
\newblock \showarticletitle{A method for file valuation in information lifecycle management}.
\newblock  (\bibinfo{year}{2007}).
\newblock


\bibitem[Wang et~al\mbox{.}(2021)]%
        {wang_data_2021}
\bibfield{author}{\bibinfo{person}{Bohong Wang}, \bibinfo{person}{Qinglai Guo}, \bibinfo{person}{Tianyu Yang}, \bibinfo{person}{Luo Xu}, {and} \bibinfo{person}{Hongbin Sun}.} \bibinfo{year}{2021}\natexlab{}.
\newblock \showarticletitle{Data valuation for decision-making with uncertainty in energy transactions: {A} case of the two-settlement market system}.
\newblock \bibinfo{journal}{\emph{Applied Energy}}  \bibinfo{volume}{288} (\bibinfo{date}{April} \bibinfo{year}{2021}), \bibinfo{pages}{116643}.
\newblock
\showISSN{0306-2619}
\urldef\tempurl%
\url{https://doi.org/10.1016/j.apenergy.2021.116643}
\showDOI{\tempurl}


\bibitem[Wang et~al\mbox{.}(2020)]%
        {wang_principled_2020}
\bibfield{author}{\bibinfo{person}{Tianhao Wang}, \bibinfo{person}{Johannes Rausch}, \bibinfo{person}{Ce Zhang}, \bibinfo{person}{Ruoxi Jia}, {and} \bibinfo{person}{Dawn Song}.} \bibinfo{year}{2020}\natexlab{}.
\newblock \showarticletitle{A principled approach to data valuation for federated learning}.
\newblock \bibinfo{journal}{\emph{Federated Learning: Privacy and Incentive}} (\bibinfo{year}{2020}), \bibinfo{pages}{153--167}.
\newblock
\newblock
\shownote{Publisher: Springer}.


\bibitem[Wang et~al\mbox{.}(2013)]%
        {wang_theoretical_2013}
\bibfield{author}{\bibinfo{person}{Yining Wang}, \bibinfo{person}{Liwei Wang}, \bibinfo{person}{Yuanzhi Li}, \bibinfo{person}{Di He}, \bibinfo{person}{Tie-Yan Liu}, {and} \bibinfo{person}{Wei Chen}.} \bibinfo{year}{2013}\natexlab{}.
\newblock \showarticletitle{A {Theoretical} {Analysis} of {NDCG} {Type} {Ranking} {Measures}}.
\newblock \bibinfo{journal}{\emph{Journal of Machine Learning Research}}  \bibinfo{volume}{30} (\bibinfo{date}{April} \bibinfo{year}{2013}).
\newblock


\bibitem[Wijnhoven et~al\mbox{.}(2014)]%
        {wijnhoven_value-based_2014}
\bibfield{author}{\bibinfo{person}{Fons Wijnhoven}, \bibinfo{person}{Chintan Amrit}, {and} \bibinfo{person}{Pim Dietz}.} \bibinfo{year}{2014}\natexlab{}.
\newblock \showarticletitle{Value-based file retention: {File} attributes as file value and information waste indicators}.
\newblock \bibinfo{journal}{\emph{Journal of Data and Information Quality (JDIQ)}} \bibinfo{volume}{4}, \bibinfo{number}{4} (\bibinfo{year}{2014}), \bibinfo{pages}{1--17}.
\newblock
\newblock
\shownote{Publisher: ACM New York, NY, USA}.


\end{thebibliography}

% \appendix

% \section{Data used in this paper}

% Table \ref{tab:ideal_data_assets_order}

\end{document}